\documentclass[conference]{IEEEtran}

\ifCLASSINFOpdf
\else
\fi
\usepackage{comment}
\usepackage{listings}
\usepackage{booktabs}
\usepackage{multirow}
\usepackage{graphicx}
\usepackage{caption}
\usepackage{subcaption}

\usepackage[flushleft]{threeparttable}

\newcommand{\PROP}{RCABench}
\newcommand{\AURORA}{AuroraFE}
\newcommand{\VULNLOC}{VulnLocFE}

\lstdefinestyle{myStyle}{
    basicstyle=\footnotesize,
}
\makeatletter % changes the catcode of @ to 11
\newcommand{\linebreakand}{%
  \end{@IEEEauthorhalign}
  \hfill\mbox{}\par
  \mbox{}\hfill\begin{@IEEEauthorhalign}
}
\makeatother % changes the catcode of @ back to 12

\begin{document}
\title{\PROP: Open Benchmarking Platform for Root Cause Analysis}

% \author{\IEEEauthorblockN{Michael Shell}
% \IEEEauthorblockA{Georgia Institute of Technology\\
% someemail@somedomain.com}
% \and
% \IEEEauthorblockN{Homer Simpson}
% \IEEEauthorblockA{Twentieth Century Fox\\
% homer@thesimpsons.com}
% \and
% \IEEEauthorblockN{James Kirk\\ and Montgomery Scott}
% \IEEEauthorblockA{Starfleet Academy\\
% someemail@somedomain.com}}

\author{
\IEEEauthorblockN{Keisuke Nishimura, Yuichi Sugiyama, Yuki Koike, \\Masaya Motoda, Tomoya Kitagawa, Toshiki Takatera, and Yuma Kurogome}\\
\IEEEauthorblockN{Ricerca Security, Inc.}\\}


%Telephone: (800) 555--1212, Fax: (888) 555--1212}


\IEEEoverridecommandlockouts
\makeatletter\def\@IEEEpubidpullup{6.5\baselineskip}\makeatother
\IEEEpubid{\parbox{\columnwidth}{
    Workshop on Binary Analysis Research (BAR) 2023 \\
    3 March 2023, San Diego, CA, USA \\
    ISBN 1-891562-84-3 \\
    https://dx.doi.org/10.14722/bar.2023.23004 \\
    www.ndss-symposium.org
}
\hspace{\columnsep}\makebox[\columnwidth]{}}

\maketitle

\begin{abstract}

Fuzzing has contributed to automatically identifying bugs and vulnerabilities in the software testing field.
Although it can efficiently generate crashing inputs, these inputs are usually analyzed manually.
Several root cause analysis (RCA) techniques have been proposed to automatically analyze the root causes of crashes to mitigate this cost.
However, outstanding challenges for realizing more elaborate RCA techniques remain unknown owing to the lack of extensive evaluation methods over existing techniques.
With this problem in mind, we developed an end-to-end benchmarking platform, \PROP{}, that can evaluate RCA techniques for various targeted programs in a detailed and comprehensive manner.
Our experiments with \PROP{} indicated that the evaluations in previous studies were not enough to fully support their claims.
Moreover, this platform can be leveraged to evaluate emerging RCA techniques by comparing them with existing techniques.

\end{abstract}


% \fi
%%\IEEEpeerreviewmaketitle

\section{Introduction}
\label{sec:introduction}

Fuzzing has contributed to automatically identifying bugs and vulnerabilities in the software testing field;
it is the process of randomly generating inputs and providing them to a program to find crashing inputs~\cite{fuzzing_state_of_the_art, fuzzing_the_art_science_survey}.
Fuzzing is simple and easy to deploy compared to other testing methods, and it can also automatically and efficiently find crashing inputs, helping an enormous number of software programs improve their quality.
In fact, OSS-Fuzz~\cite{oss_fuzz}, a popular fuzzing infrastructure, is used in more than 650 open-source software (OSS) projects and has found more than 40,500 bugs and vulnerabilities~\cite{oss-fuzz-page}.

However, crash analysis, which is required after fuzzing, is difficult and can be a bottleneck in making software testing scalable.
We must manually analyze these inputs later because fuzzers only generate crashing inputs.
The cost of analyzing the crashing inputs generated by fuzzers is very high for two reasons.
First, fuzzers sometimes generate numerous crashing inputs, the root causes of which are the same.
For example, a fuzzer generated more than 254,000 crashing inputs for 39 unique bugs in an experiment conducted in a previous study~\cite{igor}.
Second, fuzzers randomly generate crashing inputs, which means that the inputs can contain a significant amount of noise.
In other words, these inputs can include byte sequences that are not essentially related to the crash causes and hence can be removed.
During crash analysis, analysts need to determine which parts of the inputs are related to the causes, which costs significantly as the noise increases.

To reduce such costs in crash analysis, various automated techniques have been proposed in the field of \textit{triage}~\cite{fuzzing_the_art_science_survey,igor,aurora,vulnloc,default}.
Triage is the process of analyzing and reporting inputs that cause crashes~\cite{fuzzing_the_art_science_survey}.
In triage research, \textit{root cause analysis (RCA)} has attracted particular attention in recent years.
RCA, also known as localization or fault localization, is the process of identifying lines, basic blocks, or conditions related to the root cause of a crash caused by a crashing input;
it provides developers with hints about the root cause.
While there are several different RCA approaches~\cite{fl_survey_WongGLAW16, fl_survey_SouzaCK16, fl_survey_AgarwalA14}, this study focuses on statistical fault localization, which infers the program states correlated with the crash causes by contrasting the execution of crashing and non-crashing inputs.

However, RCA is relatively underdeveloped among triage topics.
As discussed in detail later, state-of-the-art RCA techniques are not infallible.
For example, in DeFault~\cite{default}, the average false positive rate was 9.2 \%.
Furthermore, RCA is not widely applied in industry, whereas deduplication, another triage technique, is integrated into major fuzzers, such as AFL~\cite{afl} and honggfuzz~\cite{honggfuzz}.
Thus, the existing RCA techniques are neither accurate nor practical enough to be widely used in the real world;
this is because RCA techniques have not been thoroughly improved owing to undiscovered outstanding challenges, that is, the problems to be solved to realize more elaborate RCA techniques.
One of the causes that make such challenges still unknown would be the lack of extensive evaluation methods over existing techniques.
We found that the following three points were unnoticed and should be considered to realize extensive evaluation methods:

    \noindent\textbf{Non-uniqueness of root cause definition}\ \
    There can be several possible patches to fix a complex bug.
    If we define (the location in the source code of) the root cause as the location that should be fixed, there can be multiple candidates for them.
    It is not obvious for evaluators to define where root causes lie, while it is certainly necessary for evaluating RCA techniques.
    Despite this vagueness, existing techniques~\cite{aurora,vulnloc,default} did not fully disclose the ground truth of their evaluations.
    This makes it difficult to reproduce the experiments in existing studies.

    \noindent\textbf{Decoupling RCA steps}\ \
        We found that the existing techniques consist of two separable steps: \textit{data augmentation} and \textit{feature extraction}.
        However, these techniques have not been evaluated separately to determine the performance of each step.

    \noindent\textbf{Variance-aware evaluation for data augmentation}\ \
        The existing techniques augment data using various fuzzing methods, particularly those that are altered for data augmentation.
        The evaluations need to consider the random nature of fuzzing.

Considering these three points,
we developed \PROP{}, an end-to-end benchmarking platform, to reveal the challenges of RCA.
We provide a detailed and comprehensive evaluation of existing techniques for various targets and find some cases where exsting techniques cannot correctly analyze.
Moreover, this platform can be used to evaluate the new RCA techniques proposed in the future by comparing them with existing techniques.

Overall, the main contributions of this work are as follows:
\begin{itemize}
    \item We present three problems in the evaluation methods of existing RCA studies.
    \item We developed \PROP{}, an open-source benchmarking platform\footnote{{https://github.com/RICSecLab/RCABench}}; it provides a more standardized evaluation and helps to summarize the outstanding challenges in RCA.
    \item Through experiments with \PROP, we identified several insights into the pitfalls of existing techniques and provided examples to motivate further research.
\end{itemize}

\section{Root Cause Analysis}
\label{sec:background}

\textit{Root cause analysis (RCA)}, also known as localization or fault localization, is a process of automatically identifying lines, basic blocks, or conditions related to the root cause of a crash;
it aids in debugging and reduces the cost of the crash analysis.
We analyzed state-of-the-art RCA techniques~\cite{aurora,vulnloc,default} and identified two separable processes that are common to all.
In this study, we refer to these as \textit{data augmentation} and \textit{feature extraction}\footnote{Although these processes were not explicitly defined in the RCA study, inspired by similar efforts in the machine learning field, we refer to them as data augmentation and feature extraction.}.
In this section, we describe these in detail.

\subsection{Data Augmentation}

Data augmentation is the process of generating new crashing and non-crashing inputs from a given crashing input;
this is the first process in RCA, and the generated inputs are used as datasets for the feature extraction.
Therefore, the quality of the dataset affects the RCA results.

The existing techniques augment inputs using various fuzzing methods
that are specially altered for data augmentation.
For example, Aurora~\cite{aurora} uses the \textit{crash exploration mode} provided by AFL~\cite{afl}, a typical coverage-guided fuzzer.
In this study, we refer to this as \textit{AFLcem}.
VulnLoc~\cite{vulnloc} proposed \textit{ConcFuzz}, a directed fuzzer for efficiently generating inputs that exercise execution paths in the neighborhood of the path taken by a given crashing input, aiming at augmentation of higher quality.
These methods use a single crashing input as the \textit{initial seed} and automatically generate crashing/non-crashing inputs by randomly mutating it.
In fuzzing, an initial seed is an input provided at the beginning of a fuzzing campaign.

\subsection{Feature Extraction}

Feature extraction analyzes the root cause using a dataset generated by data augmentation.
This process consists of two steps.
First, an analyzer records the state of the targeted program at runtime while executing it with each input in the dataset.
For example, in Aurora, executed instructions and variable values are recorded.
Next, the analyzer compares traces between the crashing and non-crashing inputs and statistically infers their differences.
This difference is indicated as the root cause.
The existing techniques estimate the lines or basic blocks related to the root cause.
Aurora also estimates predicates, that is, simple Boolean expressions that represent the conditions to be met before a crash occurs.
When actually used, analyzers do not report only the most likely root cause candidate but instead multiple candidates in descending order of the level of confidence that analyzers assign to them.
In this study, we denote VulnLoc and Aurora analyzers \textit{\VULNLOC{}} and {\textit{\AURORA{}}}, respectively.

To better illustrate this step, we take as an example CVE-2016-10094 in LibTIFF, an open-source library.
As shown in Listing~\ref{lst:CVE-2016-10094}, this vulnerability causes a heap buffer overflow owing to an off-by-one error.
Specifically, the program crashes when the variable \textsl{count} is four and the statements inside the patched \textsl{if} statement are executed.
Generally, we refer to this \textsl{if} statement as the root cause location and ``\textsl{count == 4}'' as the root cause predicate. 
If this \textsl{if} statement appears frequently in the program traces for crashing inputs and infrequently for non-crashing inputs, the statement can be identified as the root cause location.
Similarly, if there is a distinguishable difference in the value of the variable \textsl{count} between two sets of traces, the predicate can be identified.

\begin{lstlisting}[caption=Developer patch for CVE-2016-10094 in LibTIFF., label={lst:CVE-2016-10094}, language=C++]
   if(TIFFGetField(input, TIFFTAG_JPEGTABLES, 
                   &count, &jpt) != 0) {
-    if (count >= 4) {
+    if (count > 4) {
       int retTIFFReadRawTile;
       _TIFFmemcpy(buffer, jpt, count - 2);
\end{lstlisting}

\section{Challenges in RCA Evaluation}
\label{sec:motivation}

In this section, we describe three previously unconsidered points which are imperative to extensive RCA evaluations.

\subsection{Non-uniqueness of Root Cause Definition}
\label{sec:root_cause_definition}

Sometimes, there are multiple ways to fix a bug; suppose that function B triggers a bug when it processes the data produced by function A because the produced data conform to rule X, whereas B expects rule Y. 
In this case, we can make A comply with Y or B comply with X.

In such cases, if we define the root cause locations as the locations in the source code that should be fixed, there can be multiple candidates for root cause locations.
Therefore, it is difficult to correctly include all of them as the ground truth in RCA evaluations. The evaluators currently define the ground truth manually by coming up with all the possible patches, and hence, the evaluators sometimes miss some of the root cause locations and use different ground truths.

To illustrate more simply that there are multiple root cause locations for a bug, we take CVE-2017-15232 in Libjpeg as an example;
this vulnerability causes a null pointer dereference owing to the lack of a code to check for a null pointer.
There are several possible fixes for this vulnerability, as shown in Listings~\ref{lst:CVE-2017-15232} and \ref{lst:CVE-2017-15232_another}.
The first method, as shown in Listing~\ref{lst:CVE-2017-15232}, is to insert a code to check for a null pointer before the \textsl{for} statement.
Another way, as shown in Listing~\ref{lst:CVE-2017-15232_another}, is to do the same at the beginning within the \textsl{for} statement.
Thus, the root cause location is not uniquely determined, and identifying all the candidates is difficult;
this can occur frequently with bugs whose root cause is the absence of code.

The existence of multiple root cause locations makes it difficult to determine the ground truth and prevents the evaluation results from being identical.
In addition, the existing studies did not fully disclose the ground truth,
making evaluators have difficulty reproducing the existing experiments accurately.

\begin{lstlisting}[caption=Developer patch for CVE-2017-15232 in Libjpeg., label={lst:CVE-2017-15232}, language=C++]
+  if (output_buf == NULL && num_rows)
+    ERREXIT(cinfo, JERR_BAD_PARAM);
   for (row = 0; row < num_rows; row++) {
     jzero_far((void *) output_buf[row],
       (size_t) (width * sizeof(JSAMPLE)));
\end{lstlisting}
\begin{lstlisting}[caption=Another developer patch for CVE-2017-15232 in Libjpeg., label={lst:CVE-2017-15232_another}, language=C++]
   for (row = 0; row < num_rows; row++) {
+    if (output_buf == NULL)
+      ERREXIT(cinfo, JERR_BAD_PARAM);
     jzero_far((void *) output_buf[row],
       (size_t) (width * sizeof(JSAMPLE)));
\end{lstlisting}

\subsection{Decoupling Data Augmentation and Feature Extraction}

As described in Section~\ref{sec:background}, we found that state-of-the-art RCA methods~\cite{aurora,vulnloc,default} consist of two separable steps: data augmentation and feature extraction.
However, the evaluations in these previous studies did not decouple the data augmentation and feature extraction.
Evaluators should investigate the performance of each process independently because these are two separable steps.
For example, VulnLoc~\cite{vulnloc} proposed ConcFuzz as a data augmentation method but did not evaluate its relative performance by replacing it with AFLcem, an existing alternative algorithm.
In other words, it has not been fully confirmed that ConcFuzz generates datasets of higher quality than AFLcem.
Thus, the pure performance achieved by each step of proposed methods was not measured.

\subsection{Variance-aware Evaluation of Data Augmentation}

In existing studies~\cite{aurora,vulnloc,default}, evaluations did not consider the variable characteristics of data augmentation.
As described in Section~\ref{sec:background}, data augmentation generates a dataset for feature extraction using fuzzing.
Therefore, the quality of the dataset may depend on the configuration of fuzzers, such as the initial seeds and duration of a fuzzing campaign.
The existing studies have not dealt with this concern and have not been able to evaluate the impact of data augmentation on RCA results in a variance-aware manner.
Specifically, the following three variables should be considered:

\noindent\textbf{Data augmentation time}\ \
\label{sec:data_augmentation_time}
The existing studies did not evaluate RCA techniques with various values of the time spent in data augmentation.
For example, in Aurora, the data augmentation time was fixed to only one value, either 2 or 12 h, depending on targeted programs.
However, the data augmentation time can affect RCA results in multiple ways.
We can generate a dataset with a larger amount of crashing/non-crashing inputs by spending more time in fuzzing; 
this may increase the dataset diversity and improve the accuracy of feature extraction.
However, it is also plausible that overfitting occurs, similarly to data augmentation in machine learning, making the accuracy worse.

\noindent\textbf{Initial seed}\ \
\label{sec:initial_seed}
The existing studies prepared only one specific crashing input as an initial seed for fuzzing in data augmentation.
For example, in VulnLoc, the initial seed is the input used as a proof-of-concept when reporting vulnerabilities.
In fuzzing, the difference in the initial seeds is known to affect performance, such as coverage and bug finding~\cite{seed_selection_for_successful_fuzzing, explainable_fuzzer_evaluation}.
Data augmentation using fuzzing may also affect the accuracy of feature extraction; therefore, the evaluator should prepare several initial seeds.
In addition, the existing studies have not focused on the characteristics of initial seeds.
For example, the initial seed generated by a fuzzer tends to be noisier and more complex than that generated by an analyst manually.

\noindent\textbf{Fuzzing randomness}\ \
\label{sec:statistical_evaluation}
The existing studies have not considered the randomness of fuzzing in data augmentation.
They evaluated each method using only a dataset from a single fuzzing run.
However, fuzzing is a highly stochastic process.
Hence, the generated dataset changes with each run, which can affect the results of RCA.
RCA techniques must be evaluated multiple times to address this problem.
In fuzzing studies, it has already been standard practice to run fuzzers multiple times and evaluate the results statistically if possible~\cite{ICSE22_fuzzing_evaluation}.
The same approach is required in RCA studies.

\section{Proposal: \PROP{}}
\label{sec:proposal}
We propose \PROP{}, an end-to-end benchmarking platform that can run RCA techniques on selected bugs and check whether their results match the predefined locations of root causes\footnote{Some RCA techniques (e.g., VulnLoc and Aurora) indicate the candidates for root cause locations as pairs of addresses of an assembly instruction and their corresponding source line numbers.
\PROP{} uses line numbers for the check because the addresses of instructions are too fine-grained to decide whether the address is a root cause location.
}.
The design of the \PROP{} was motivated by the insights described in Section \ref{sec:motivation}.
For each RCA technique, the data augmentation and feature extraction steps were decoupled, which enabled the comparison and evaluation of the augmentation and extraction methods separately.

Currently, \PROP{} supports two augmentation methods and two extraction methods.
The available augmentation methods are \textit{AFLcem}, used in Aurora~\cite{aurora},
and \textit{ConcFuzz}\footnote{For ConcFuzz, the time spent in saving its internal data at the end is not included in the augmentation time.}, proposed in VulnLoc~\cite{vulnloc}.
The available extraction methods are the \AURORA{} and \VULNLOC{}.
We decoupled the augmentation and extraction steps and abstracted their interfaces so that each augmentation method could be connected to each extraction method interchangeably since the original implementations of \AURORA{} and \VULNLOC{} are incompatible with ConcFuzz and AFLcem, respectively.
Consequently, \PROP{} can evaluate previously untested combinations of AFLcem $\times$ \VULNLOC{} and ConcFuzz $\times$ \AURORA{}. 
Note that we used only the root cause locations inferred by \AURORA{} to compare the performance of the techniques.
Supporting and evaluating the root cause predicates included in the outputs of \AURORA{} are left for future work.

\PROP{} provides multiple popular real-world programs containing actual bugs and vulnerabilities as targets of RCA. Currently, seven targets have been prepared, all of which were used in the evaluations of Aurora and VulnLoc~\cite{aurora, vulnloc}.
We show the lists and summary of their root causes in Table~\ref{tab:root_cause_info}.
Our criteria for selecting targets are the availability of the source code and the diversity of root and crash causes.
As discussed in the previous section, the selection of root cause locations can vary and be biased.
Therefore, we first registered to \PROP{} several reasonable candidates for root cause location as ground truth for each target.
For stable re-evaluation, \PROP{} publicly exposes these root cause locations, along with their brief explanations.
\PROP{} also includes one or more initial seeds for each target to support augmentation methods that require a crashing input as an initial seed.
For targets with multiple seeds available, we selected a crashing input used in bug disclosure or explanation to the developers as the baseline.

\section{Benchmark Results}
\label{sec:evaluation}
This section describes the results of the proposed benchmark \PROP{}.
Through benchmarking, we answered the following questions:
\begin{itemize}
    \item \textbf{RQ1:} Which RCA techniques can perform accurate analysis on each bug?
    \item \textbf{RQ2:} Does the increase in data augmentation time improve accuracy?
    \item \textbf{RQ3:} Do initial seeds affect accuracy?
    \item \textbf{RQ4:} Does the randomness of data augmentation affect accuracy?
\end{itemize}

All results shown here were obtained on a 256-CPU (AMD EPYC 7742) machine with 2TB memory and Ubuntu 20.04 operating system.
To investigate the relationship between the data augmentation time and the accuracy of RCA techniques,
we ran the data augmentation process up to an imposed time limit and, in each of the first 5, 15, 30, and 45 minutes and every hour thereafter during the execution,
\PROP{} saved the dataset produced by the data augmentation at that time and analyzed root cause with the dataset.
For Targets \#1-4,6, we set the time limit to 4h, and for Targets \#5,7, we extended it to 12h in accordance with the evaluation of Aurora~\cite{aurora}.

\begin{table}[tb]
\centering
 \begin{threeparttable}
    \caption{Results of four RCA techniques in different data augmentation times.}
    \label{tab:overview_result}
    \begin{tabular}{cc|c|cccc}
        \toprule
         & Program & D.A. Time & 
         A $\times$ A &  C $\times$ A & A $\times$ V & C $\times$ V \\
        \midrule
       \multirow{3}{*}{\#1} & \multirow{3}{*}{LibTIFF} 
                  & 15 m & 15 &  9 & 2 & 13 \\
               && 2 h & 9 &  33 & 2 & 12\\
               && 4 h & 9 &  47 & 2 & 12\\
        \midrule
        \multirow{3}{*}{\#2} & \multirow{3}{*}{Libjpeg} & 15 m & -- & -- & 23 & 32\\
                                            && 2 h    & -- & 15 & 12 & 23\\
                                            && 4 h    & -- & 14 & 12 & 17\\
        \midrule
        \multirow{3}{*}{\#3} & \multirow{3}{*}{Libjpeg} & 15 m & 16 & -- & 2 & 1\\
                            && 2 h & 7 & -- & 2 & 1\\
                            && 4 h & 6  & -- & 2 & 1\\
        \midrule
        \multirow{3}{*}{\#4} & \multirow{3}{*}{Libxml2} & 15 m & 28 & -- & 19 & 16 \\
                                               && 2 h & 28 & -- & 23 & 16 \\
                                               && 4 h & 27  & -- & 19 & 17 \\
        \midrule
        \multirow{3}{*}{\#5} & \multirow{3}{*}{mruby} & 15 m & 41 & 105 & 59 & 11\\
              && 4 h   & 30 & 60 & -- & 45\\
              && 12 h  & 31 & 60 & -- & 45\\
        \midrule
        \multirow{3}{*}{\#6} & \multirow{3}{*}{readelf} & 15 m & 1 & 4 & 4 & 4\\
                                            && 2 h & 1 & 1 & 4 & 4\\
                                            && 4 h & 1 & 1 & 4 & 4\\
        \midrule
        \multirow{3}{*}{\#7} &  \multirow{3}{*}{Lua} &       15 m & --   & -- & 1 & 1\\
                               && 4 h & --  & N/A  & -- & N/A\\
                               && 12 h & 32 & N/A  & -- & N/A\\
        \bottomrule
    \end{tabular}
    \begin{tablenotes}
      \item ``C $\times$ A'' means ConcFuzz $\times$ \AURORA{} and ``A $\times$ V'' means AFLcem $\times$ \VULNLOC{}.
      \item  ``N/A'': No data were obtained.
             ``--'':  The root cause location did not appear in the candidates reported by an RCA technique.
    \end{tablenotes}
  \end{threeparttable}
\end{table}

\subsection{Which RCA techniques can perform accurate analysis on each bug? (RQ1)}

Table~\ref{tab:overview_result} presents the overall comparisons of RCA techniques for each target.
The numbers in the table indicate the rank of the correct answer (actual root cause location) among the location candidates reported by each RCA technique, ordered by the level of confidence assigned by the technique.
A lower rank indicates that the RCA technique can infer the root cause location more accurately (``1'' is the best score).
In this experiment, we set VulnLocFE to report up to Top-200 candidates, in accordance with the original paper.
``--'' indicates that the correct answer was not included in the candidates produced by the RCA technique.
``N/A'' indicates that no data could be obtained because the technique tried to produce quite huge files or took a long time for file I/O, which were impossible to handle with our limited machine resources.

Table~\ref{tab:overview_result} indicates that no technique can predict the root cause locations with high accuracy for \textit{all} targets: while ConcFuzz $\times$ \VULNLOC{} successfully inferred the correct location of the root cause in Target \#3 with the highest accuracy, 
its predictions for Targets \#6 were less accurate than AFLcem $\times$ \AURORA{}.
Our newly tested combination, AFLcem $\times$ \VULNLOC{}, outperformed the existing methods for Targets \#1 and \#2.
However, they failed to find the root causes of some other targets.
This result implies that the characteristics of the targets, which can be analyzed with high accuracy, would be different for each RCA technique. This up-and-down situation depending on the targeted programs is similar to fuzzer benchmarking, as seen in some results~\cite{unifuzz, fuzzbench_reports}.

\textbf{Answer:} The technique that gave the highest rank to the correct root cause was different for each bug, and there was no universal technique that was most accurate.

\subsection{Does the increase in data augmentation time improve accuracy? (RQ2)}

\begin{figure}[tb]
    \vspace*{0cm}
    \begin{subfigure}[b]{0.47\columnwidth}
        \includegraphics[width=\columnwidth]{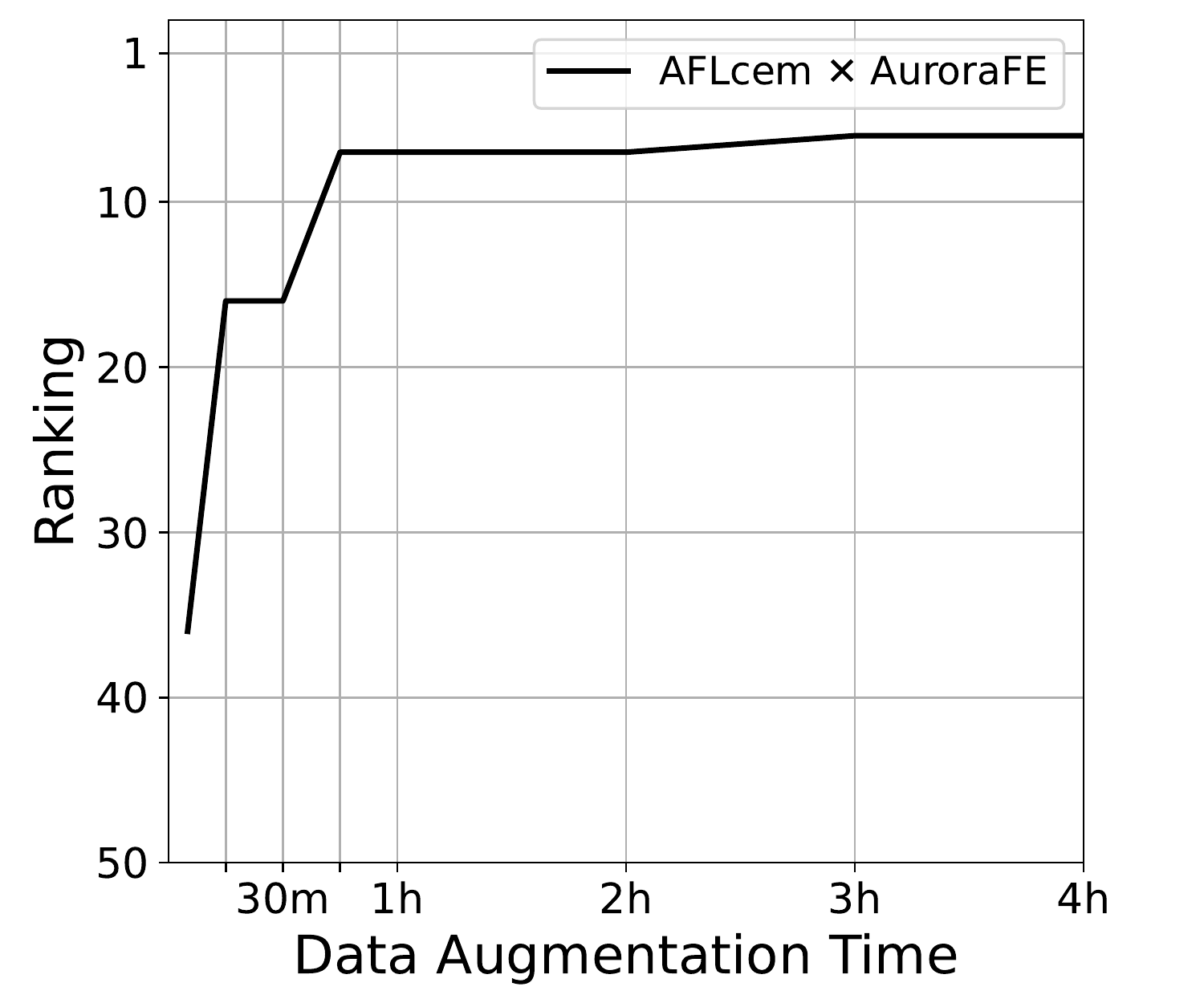}
        \subcaption{Target \#3}\label{fig:impr}
    \end{subfigure}
    \begin{subfigure}[b]{0.47\columnwidth}
        \includegraphics[width=\columnwidth]{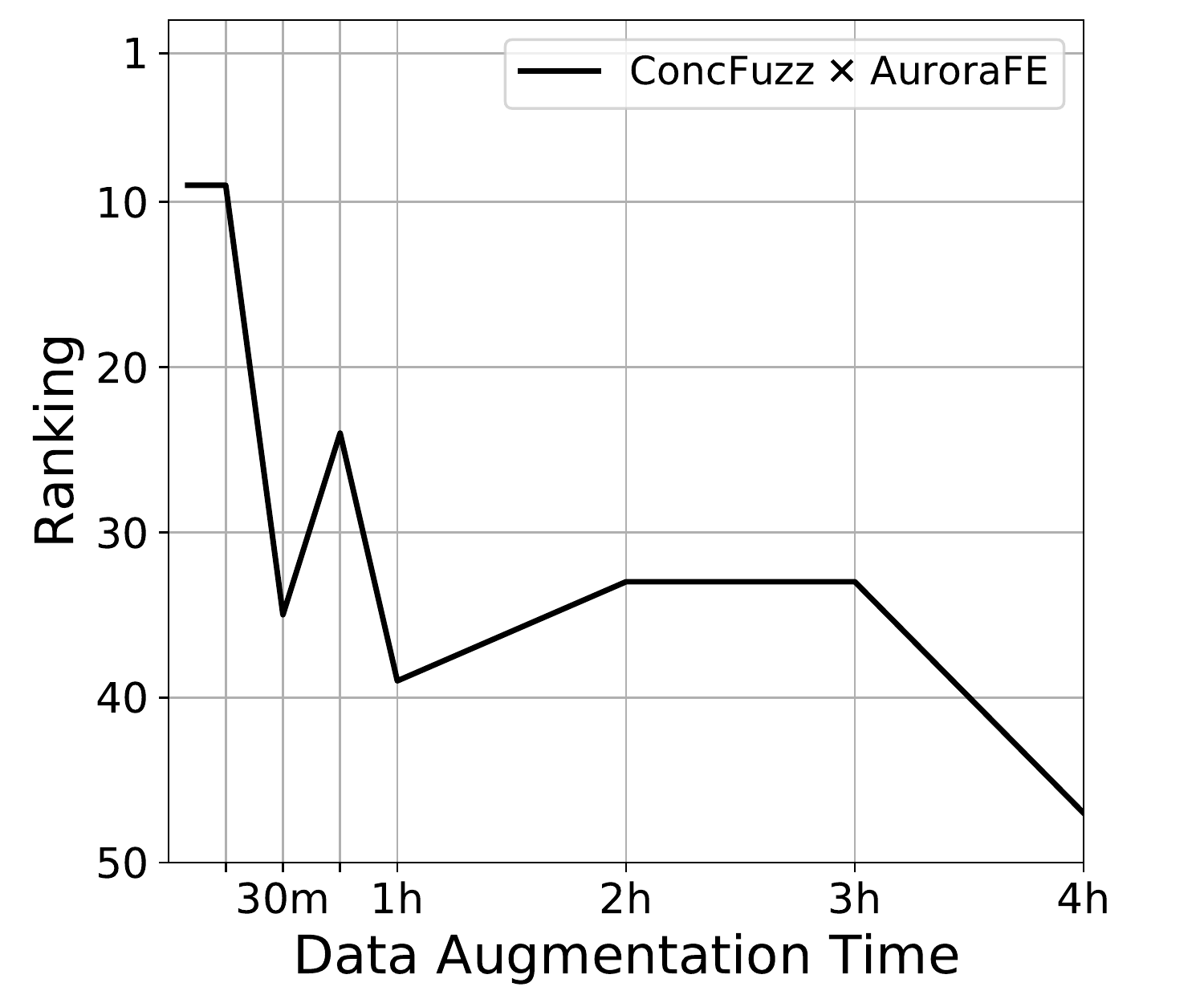}
        \subcaption{Target \#1}\label{fig:degr}
    \end{subfigure} 
    \vspace*{-0.3cm}
    \caption{Accuracy vs. data augmentation time.
    }
    \vspace*{-0.3cm}
    \label{fig:time}
\end{figure}

Next, we compared the results of each technique for each target with different data augmentation times.
Table~\ref{tab:overview_result} lists the results in three data augmentation times (15m, 2h, 4h for Targets \#1-4,6 and 15m, 4h, 12h for Targets \#5,7) for each technique and target pair.
We also show two detailed examples of how accuracy changes with time in Figure \ref{fig:time}.
In 19 cases, out of the 26 results excluding ``N/A'',
the accuracy remained the same (e.g., Target \#6 except ConcFuzz $\times$ \AURORA{}) or improved as the data augmentation time increased (e.g., AFLcem $\times$ \AURORA{} on Target \#3 in Figure \ref{fig:impr}).

However, the increase in data augmentation time worsened the accuracy in some pairs of RCA techniques and targets, such as ConcFuzz $\times$ \VULNLOC{} on Target \#5.
Figure~\ref{fig:degr} shows an example of the deteriorating trend of ConcFuzz $\times$ \AURORA{}  on Target \#1.
In this example, the highest accuracy was achieved up to 15m,
 and its accuracy declined thereafter.

\textbf{Answer:} 
While the accuracy improved or did not change over time in many cases, there were a few cases in which the accuracy was degraded.

\begin{figure}[tb]
    \vspace*{0cm}
    \begin{subfigure}[b]{0.49\columnwidth}
         \includegraphics[width=\columnwidth]{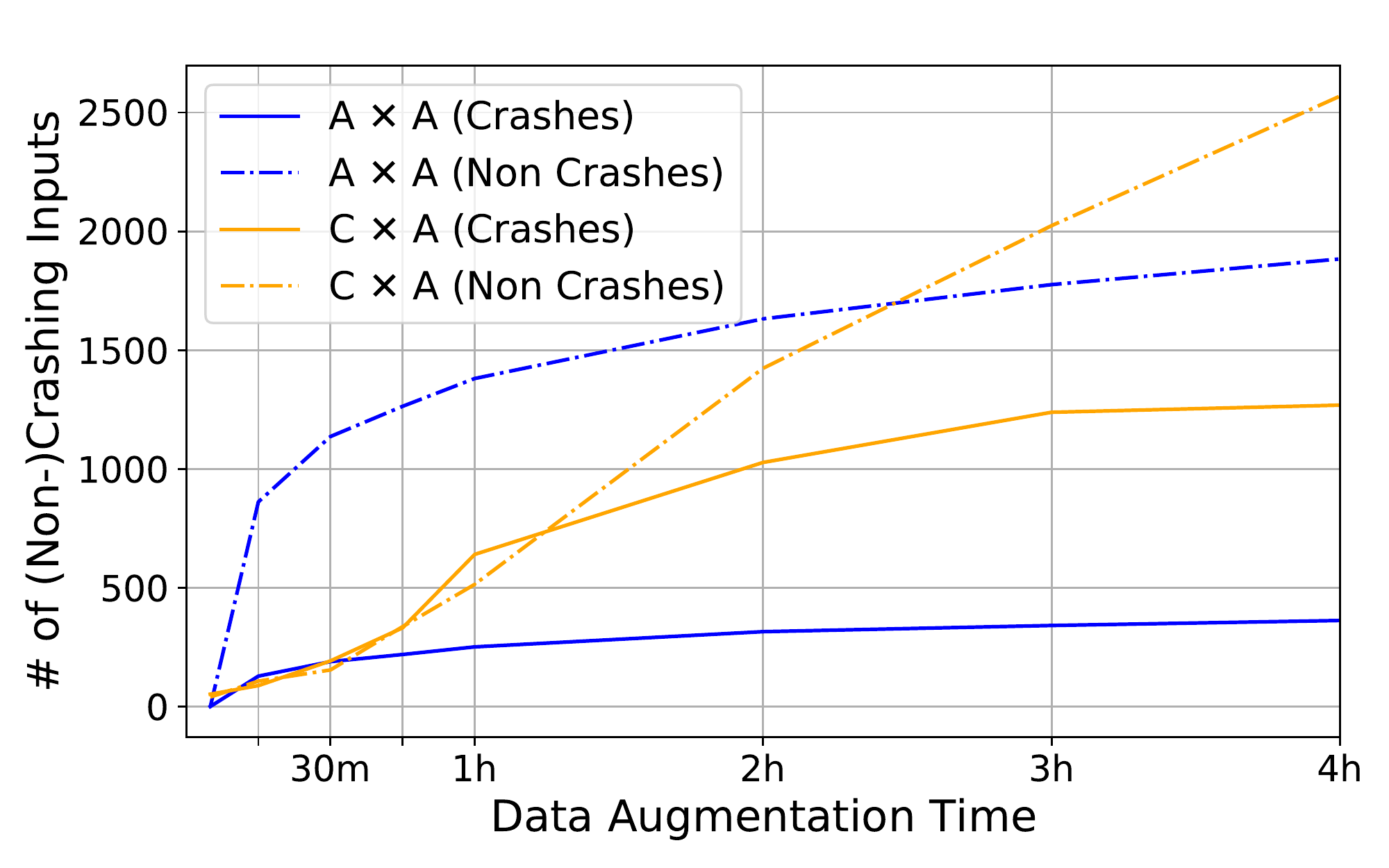}
         \subcaption{Target \#1}\label{fig:trace_libtiff}
     \end{subfigure}
    \begin{subfigure}[b]{0.49\columnwidth}
         \includegraphics[width=\columnwidth]{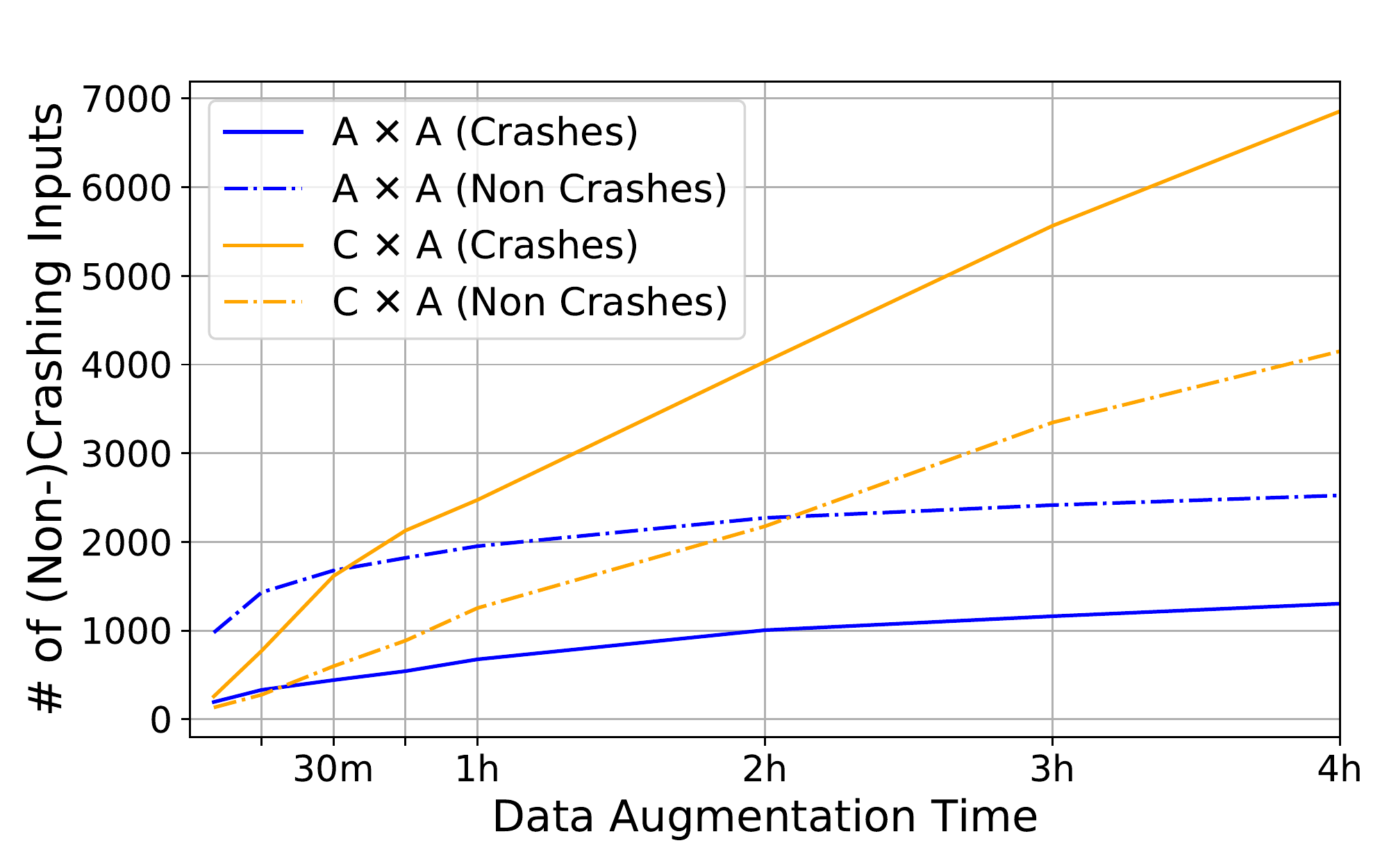}
         \subcaption{Target \#6}\label{fig:readelf}
    \end{subfigure}
    \vspace*{-0.2cm}
    \caption{The number of generated inputs over time. 
    A $\times$ A, and  C $\times$ A denote AFLcem $\times$ \AURORA{}, and ConcFuzz $\times$ \AURORA{}, respectively.
    }
    \vspace*{-0.5cm}
    \label{fig:traces}
\end{figure}

Thus, sometimes data augmentation time eventually affects the accuracy. This fact indicates that somehow data augmentation time affects the quality of datasets produced by data augmentation.
To analyze how it affects the quality, we inspected how the number and ratio of samples (i.e., crashing/non-crashing inputs) in a dataset changes as data augmentation time increases. Figure~\ref{fig:traces} plots the number of samples versus data augmentation time for Target \#1 and \#6. 

When looking at the ratio of samples produced by ConcFuzz $\times$ \AURORA{} in Figure~\ref{fig:trace_libtiff}, we see that the number of non-crashing inputs starts exceeding that of crashing inputs considerably in one hour. This would force feature extraction methods to find out root cause locations with imbalanced datasets of crashing/non-crashing inputs, which is very similar to a situation called \textit{imbalanced data classification} in the machine learning field~\cite{imbalanced_data}. Generally, imbalanced data can cause poor accuracy in these classifying tasks. Actually, in Table~\ref{tab:overview_result}, the accuracy of ConcFuzz $\times$ \AURORA{} perceptibly decreases as the ratio gets imbalanced.

Another implicative fact is that the numbers of samples for Target \#6 shown in Figure~\ref{fig:readelf} are large from the beginning, compared to those of Target \#1. Perhaps, this would have made all the RCA techniques achieve the high accuracy in Target \#6 even in 15m, considering that a large number of samples usually leads to high accuracy in classifying tasks.

\subsection{ Do initial seeds affect accuracy? (RQ3) }
To answer RQ3, we investigated whether the accuracy changed depending on the initial seed.
For this purpose, we first ran AFLcem against Target \#4 and \#6 to produce various crashing inputs. 
Then, for each target, we randomly selected two different-length inputs from the produced inputs as initial seeds and evaluated the RCA techniques with them.
We selected these two targets because we could find the crashing inputs that were much smaller or larger than the original crashing input. 
Thus, we believe that the newly produced initial seeds were very different from the original ones, in terms of seed size and the method of producing them. Note that most of the original seeds were created manually, which may make a significant difference between the original and new seeds with regard to whether noise exists in them, as described in Section~\ref{sec:introduction}.
\begin{figure}[tb]
    \vspace*{0cm}
  \begin{subfigure}[b]{0.47\columnwidth}
         \includegraphics[width=\columnwidth]{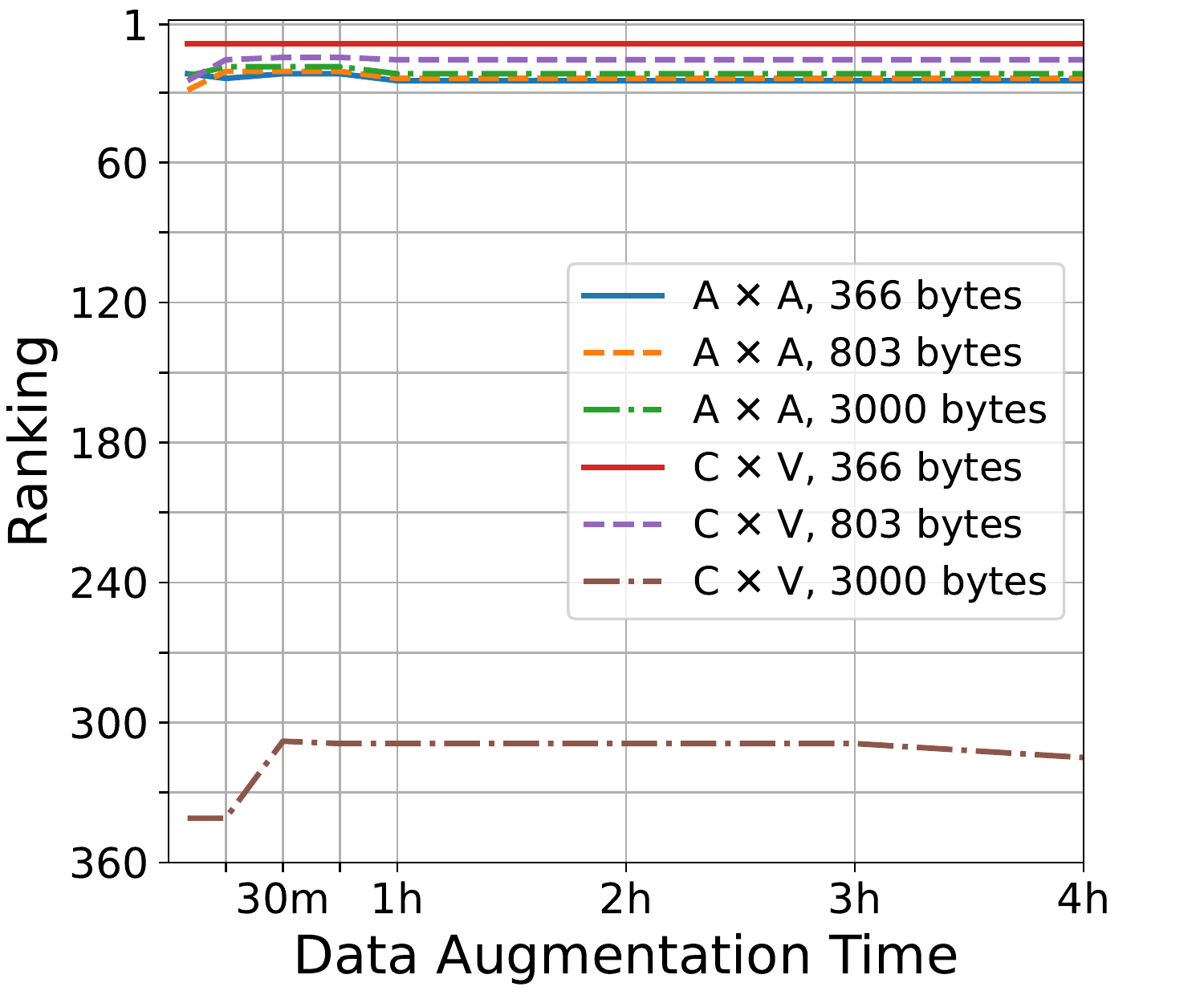}
         \subcaption{Target \#4}\label{fig:seed_libxml}
     \end{subfigure}
    \begin{subfigure}[b]{0.47\columnwidth}
         \includegraphics[width=\columnwidth]{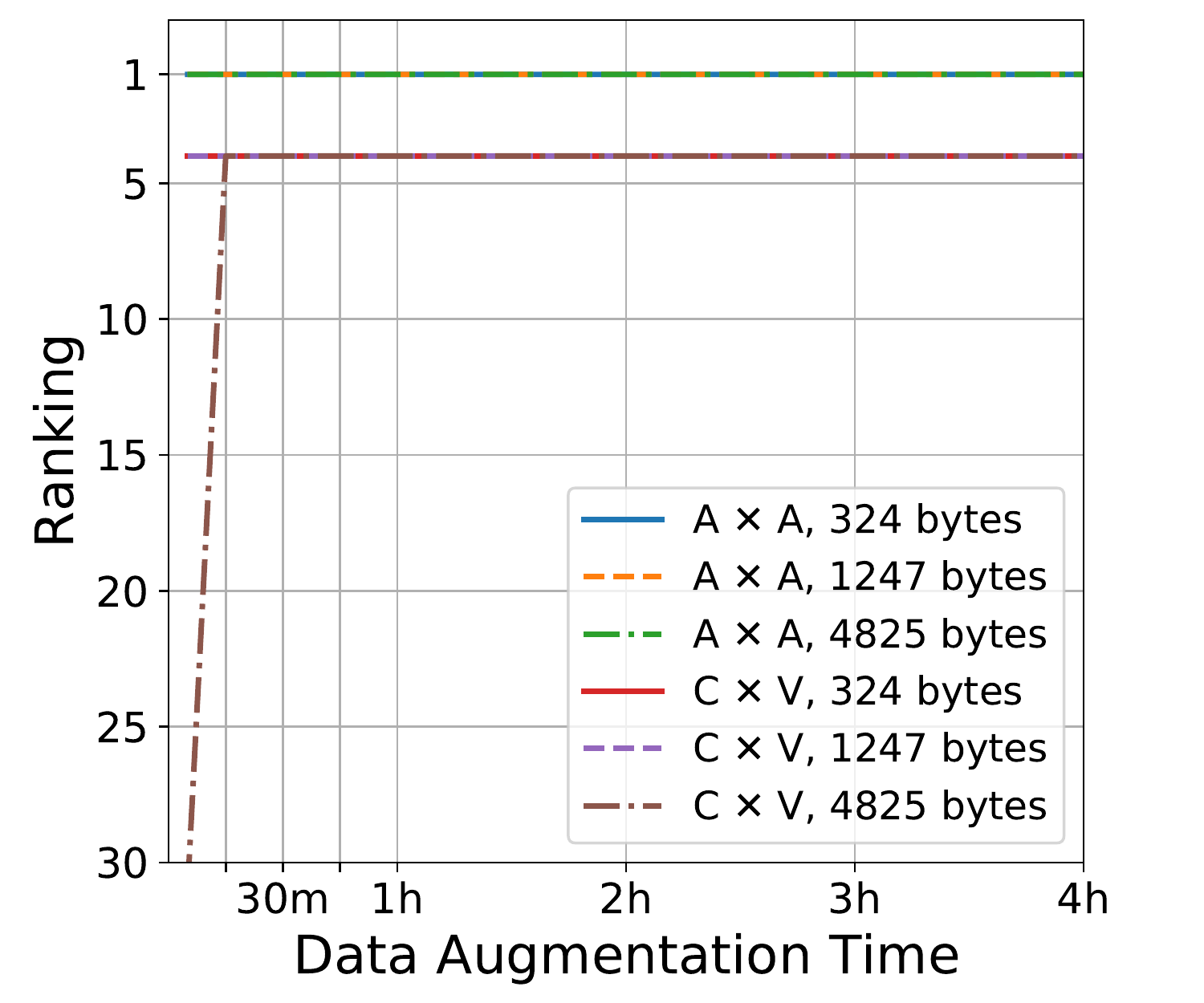}
         \subcaption{Target \#6}\label{fig:seed_readelf}
     \end{subfigure}
    \vspace*{-0.3cm}
    \caption{The transitions of the accuracy over time for different initial seeds.
    A $\times$ A, and  C $\times$ V denote AFLcem $\times$ \AURORA{} and ConcFuzz $\times$ \VULNLOC{}, respectively. In Target \#4 and \#6, the 803-byte and 324-byte seeds are their original seed, respectively.
    }
    \vspace*{-0.5cm}
    \label{fig:seeds}
\end{figure}

Figure~\ref{fig:seeds} shows accuracy versus data augmentation time for different initial seeds.
While, in Target \#6, the accuracies are little affected by the difference of initial seeds in both AFLcem $\times$ \AURORA{} and ConcFuzz $\times$ \VULNLOC{}, the two added initial seeds of Target \#4 affected the accuracy of ConcFuzz $\times$ \VULNLOC{}, and one of them had a significant impact in particular.
This result is consistent with the fact that the performance of a fuzzer can be affected by the initial seeds~\cite{seed_selection_for_successful_fuzzing, explainable_fuzzer_evaluation}.

\textbf{Answer:} 
The difference in initial seeds sometimes affects accuracy.
This implies that evaluators should make the initial seeds public to avoid cherry-picking and for reproducibility.

\subsection{Does the randomness of data augmentation affect accuracy? (RQ4)}
We observed the randomness effect on the accuracy by evaluating the techniques five times.
For ConcFuzz, different seeds of its random number generator were set in each trial.

\begin{figure}[tb]
    \vspace*{0cm}
  \begin{subfigure}[b]{0.47\columnwidth}
         \includegraphics[width=\columnwidth]{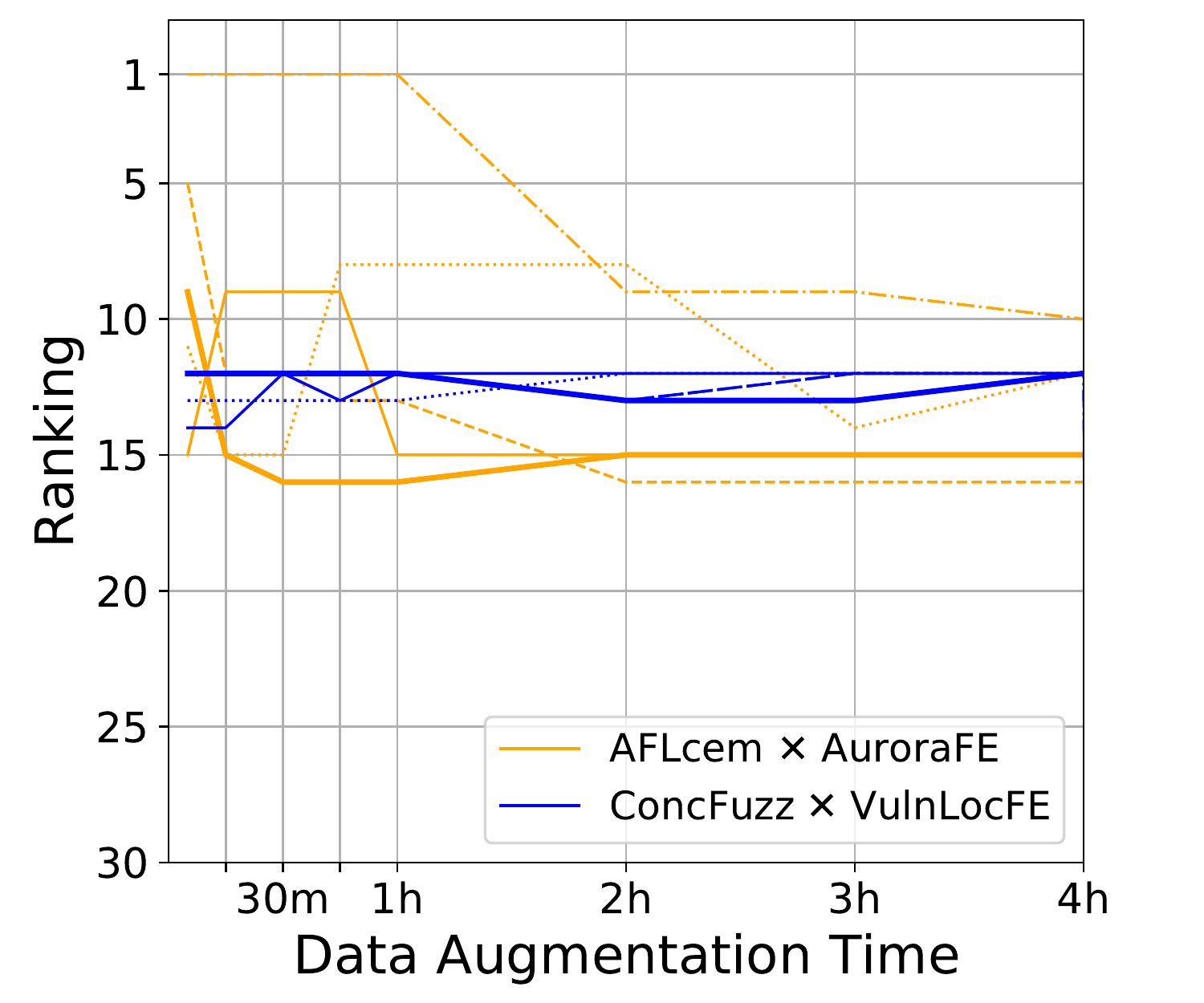}
         \subcaption{Target \#1}\label{fig:rand_libtiff}
     \end{subfigure}
  \begin{subfigure}[b]{0.47\columnwidth}
         \includegraphics[width=\columnwidth]{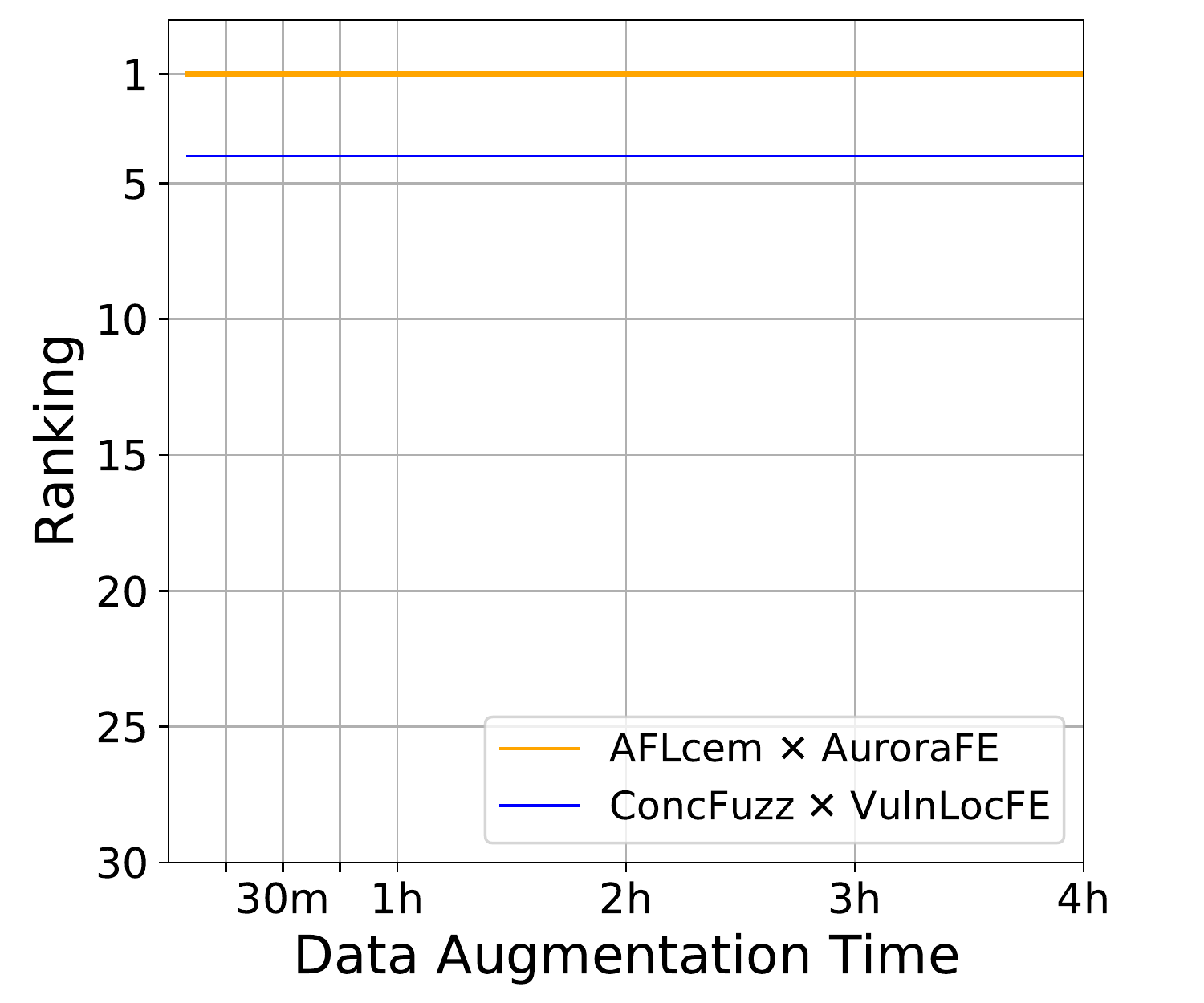}
         \subcaption{Target \#6}\label{fig:rand_readelf}
     \end{subfigure}
    \vspace*{-0.3cm}
    \caption{ The effect of the randomness in data augmentation on the accuracy. 
    Five attempts were performed.
    } 
    \vspace*{-0.5cm}
    \label{fig:random}
\end{figure}

Consequently, we observed some non-negligible variances as predicted in Section~\ref{sec:motivation}, while the accuracy was very stable in some targets.
Figure~\ref{fig:random} shows the results of AFLcem $\times$ \AURORA{} and ConcFuzz $\times$ \VULNLOC{} in Target \#1 and \#6.
While both RCA techniques in Target \#6 and ConcFuzz $\times$ \VULNLOC{} in Target \#1 had little divergence in their accuracy,
AFLcem $\times$ \AURORA{} showed significant divergence in the accuracy in Target \#1.
Specifically, one of the five trials outperformed the others.
If one cherry-picked only this trial,
it could be concluded that AFLcem $\times$ \AURORA{} was more accurate than ConcFuzz $\times$ \VULNLOC{},
although the two techniques achieved similar accuracies on average;
this suggests that it is important to evaluate RCA techniques multiple times and, if possible, perform statistical analysis of the results.

\textbf{Answer:} For some combinations of techniques and targets, randomness in data augmentation leads to non-negligible variances in accuracy.
This result suggests that experiments to evaluate RCA techniques should be conducted multiple times
to reduce the effect of randomness as much as possible.

\section{Discussion and Future Work}

\subsection{Threats to Validity}
Although the evaluations on \PROP{} and our research questions provided thought-provoking claims in Section~\ref{sec:evaluation}, we admit that two major threats may spoil some of the claims. 

The first one is the non-uniqueness of root cause definition.
As previously noted, we have been aware that multiple root cause locations can be the ground truth, and mitigated risk by making our definition public and upgradable.
However, this is just a temporary countermeasure in the sense that particular techniques can be underestimated still; it is possible that some techniques report different valid root cause locations than our definition while the others report ours.
A more robust and better definition and evaluation method of accuracy is desired.

The second one is fuzzing randomness.
While RQ4 revealed that it undoubtedly threatens the evaluation validity in previous studies, it also threatens the validity of our evaluation, especially for RQ1 and RQ3 (note that RQ2 should be cared about even within one trial, and hence its claim is valid in that sense regardless of the threats).
We acknowledge that our results do not have a statistical significance due to our limited computational resources and should be carefully reviewed by others.
Nevertheless, it is sure that the effectiveness and superiority of the existing techniques are at least not so obvious as the evaluations in the existing studies claimed.
Moreover, the lack of a statistical significance can be eventually resolved in the future since we release \PROP{} as an OSS platform and other researchers are also able to take benchmarks.

While the above two points threaten the internal validity, the external validity is another concern because seven programs are not enough to fully understand the behavior and performance of RCA techniques against a wide variety of targeted programs.
For preparing more target programs with diverse root causes, the programs constituting benchmarks known in the fuzzing field ~\cite{lava,magma,unifuzz,fuzzbench} would be reasonable candidates.
In particular, Magma~\cite{magma} and FuzzBench ~\cite{fuzzbench} provide a suite of programs that contain bugs found in the real world and are widely used.
Evaluating techniques with these programs would further clarify their practical effectiveness.

\subsection{Possible Improvement of Existing Techniques}

In our experiments, we found some cases where RCA techniques failed to analyze the root causes with high accuracy, owing to their nature. A striking example is that ConcFuzz $\times$ \AURORA{} ranked the root cause locations of Target \#1 (CVE-2016-10094) very low in RQ1 at data augmentation times of 2h and 4h; this is probably because the ideal difference that should exist between its generated crashing/non-crashing inputs is whether or not a certain variable is at a certain value, while ConcFuzz focuses on control flow and the generated inputs did not have enough diversity of values.
This suggests the possibility of further improvement of data augmentation methods by considering features other than the control flow.

Another concern about the existing techniques is that implementation designs differ among them.
For example, to trace a program, Aurora~\cite{aurora} uses Intel PIN~\cite{intel_pin} and VulnLoc uses DynamoRIO~\cite{dynamorio}.
In addition, the data augmentation method of VulnLoc is written in the programming language Python, whereas that of Aurora in C/C++.
Thus, there is a possibility that they have different performances owing to their implementation methods.
If VulnLoc were written in C/C++, VulnLoc would be able to run faster.
In the fuzzing field, some frameworks have already been proposed so that different algorithms can be implemented in a uniform way to solve such a problem~\cite{libafl,fuzzuf}.
For example, LibAFL~\cite{libafl} is a framework for building fuzzers in a modular manner.
LibAFL can reduce the cost of combining multiple fuzzing algorithms into a single fuzzer and can fairly and objectively evaluate the algorithms within the common implementation.
If a similar modular framework is presented for RCA, researchers would be able to take fairer evaluations.
Also, it would allow them to implement and evaluate a new algorithm more easily.

\section{Conclusion}

Although fuzzing is a mature method for automatically finding bugs, \textit{root cause analysis (RCA)} techniques for discovered bugs are not full-grown.
One of its causes is that the environment for a comprehensive evaluation of existing RCA techniques was inadequate, making it difficult to discover the outstanding problems.
Therefore, we developed a benchmark platform, \PROP{}, for automatic and extensive evaluation.
Our experiments indicated that the evaluations in previous studies were not enough to fully support their claims and found some cases where the representative techniques failed to analyze with high accuracy.

We believe that this initiative fosters future RCA research by assisting researchers to propose and evaluate emerging RCA techniques, as this study gives a glimpse of it.
To shed light on and help resolve the hidden challenges of RCA, we would like to continue adding various targets and techniques and making \PROP{} a more insightful platform.

\section*{Acknowledgment}
We would like to thank the anonymous reviewers for their helpful feedback.
We also gratefully acknowledge the authors of Aurora and VulnLoc,
who made their implementations and experiment configurations publically available.
This work was supported by the Acquisition, Technology \& Logistics Agency (ATLA) under the Innovative Science and Technology Initiative for Security 2020 (JPJ004596).



%\bibliographystyle{IEEEtranS}


% \end{thebibliography}

\bibliographystyle{IEEEtran}
\bibliography{ref.bib}

\clearpage
\section*{Appendix}

\begin{table}[h]
  \centering
  \caption{Details of targeted vulnerabilities.}
  \label{tab:root_cause_info}
    \begin{tabular}{ccccc}
        \toprule
         & Program & CVE ID & Root Cause & Crash Cause \\
        \midrule
       \#1 & LibTIFF & CVE-2016-10094 & off-by-one error & heap buffer overflow \\
        \midrule
        \#2 & Libjpeg & CVE-2018-19664 & incomplete check & heap buffer overflow \\
        \midrule
        \#3 & Libjpeg & CVE-2017-15232 & missing check & null pointer dereference \\
        \midrule
        \#4 & Libxml2 & CVE-2017-5969 & incomplete check & null pointer dereference \\
        \midrule
        \#5 & mruby & None & missing check & type confusion \\
        \midrule
        \#6 & readelf & CVE-2019-9077 & missing check & heap buffer overflow \\
        \midrule
        \#7 &  Lua & CVE-2019-6706 & missing check & use-after-free \\
        \bottomrule
    \end{tabular}
    \begin{tablenotes}
      \item Target \#5 was not assigned a CVE ID but was assigned ID 185041 in the HackerOne platform.
    \end{tablenotes}
\end{table}

\end{document}